\documentclass[10pt]{article}

\pdfoutput=1
\usepackage[pdftex]{color}
\usepackage{amssymb}
\usepackage{amsthm} 
\usepackage{amsmath}
\usepackage{latexsym}
\usepackage{amscd}
\usepackage{graphicx}
\usepackage[pdftex, colorlinks=true, citecolor=green]{hyperref}
\usepackage{lscape}
\usepackage{multirow}
\usepackage{wrapfig}

\newcommand{\ben}{\begin{equation}}     
\newcommand{\eeqn}{\end{equation}}
\newcommand{\bey}{\begin{eqnarray}}
\newcommand{\eey}{\end{eqnarray}}

\newtheorem{thm}{Theorem}[section]

\begin{document}

\noindent {\Large
\textbf{Modeling the effects of adherence to vaccination and health\\ protocols in epidemic dynamics by means of an SIR model}
}
\\\\
\indent Jasmin Nunuvero$^1$, Angelique Santiago$^1$, Moshe Cohen$^1$, Anca R\v{a}dulescu$^{*,}\footnote{Associate Professor, Department of Mathematics, State University of New York at New Paltz; New York, USA; Phone: (845) 257-3532; Email: radulesa@newpaltz.edu}$
\\
\indent $^1$ Department of Mathematics, SUNY New Paltz, NY 12561

\begin{abstract}
\noindent Susceptible-Infected-Recovered (SIR) models have been used for decades to understand epidemic outbreak dynamics. We develop an SIR model specifically designed to study the effects of population behavior with respect to health and vaccination protocols in a generic epidemic. Through a collection of parameters, our model includes the traditional SIR components: population birth, death, infection, recovery and vaccination rates, as well as limited immunity. 

\vspace{1mm}
\noindent We first use this simple setup to compare the effects of two vaccination schemes, one in which people are vaccinated at a rate proportional with the population, and one in which vaccines are administered to a fraction of the susceptible people (both of which are know strategies in real life epidemics). We then expand on the model and the analysis by investigating how these two vaccination schemes hold under two scenarios of population behavior: one in which people abide by health protocols and work towards diminishing transmission when infection is high; one in which people relax health protocols when infection is high. 

\vspace{1mm}
\noindent We illustrate these two aspects (vaccination and adherence to health protocols) act together to control the epidemic outbreak. While it is ideal that the tow components act jointly, we also show that tight observance of health protocols may diminish the need for vaccination in the effort to clear or mitigate the outbreak. Conversely, an efficient vaccination strategy can compensate for some degree of laxity in people's behavior.
\end{abstract}

\section{Introduction}

Improved strategies for epidemic outbreak mitigation and control have been a matter of utmost interest in healthcare, and they have become even more central in light of the COVID 19 pandemic. Whether looking at a new outbreak or an endemic condition, it is often the case that many of the specific epidemic measures (such as incubation period, $R_0$ number, recovery time, immunity prevalence) are either already known or can be obtained relatively fast from clinical and epidemiological data. These parameters altogether represent the characteristics of the epidemic. Clinical alternatives typically look for medical treatments that would alter these measures in a way that leads to a better epidemic outcome. Examples of improving the outcome include minimizing the number of people infected, the time spent being sick, or the casualty rate, or simply shortening the outbreak. 

During an outbreak that poses high concerns, healthcare systems search, concomitantly with clinical answers, for adequate strategies of epidemic control. For successful implementation of such protocols, it is invaluable to gain \emph{a priori} knowledge of the timing and extent that would lead to optimal control of the outbreak. One of the important strategies to obtain such predictive knowledge is that of creating mathematical models of the outbreak at hand. Built based on the known epidemic measures, but also incorporating the control key parameters (e.g., health and vaccination protocols), the model can be then used to explore the effects of tuning these key parameters on the long-term prognosis of the epidemic system.

More specifically, factors that are hypothesized to affect the system's evolution and prognosis can be introduced as constant parameters or as state-dependent functions. One of the attributes of such simple models is to allow tractable analytical or numerical computations that express directly how the behavior of the system depends on these parameters. Successful models are often able to answer rather ambitious mathematical questions and generate testable predictions which can be further used for informed decisions on how to control the system.

\section{Modeling methods}

Historically, compartmental models have been used successfully to study  and predict epidemic dynamics. They rely on the idea of dividing the population into epidemiology-based compartments and writing equations representing the flow between them. In such models, the compartments are viewed as black boxes, in the sense that no specific detail aims at describing their inner dynamics. However, the communication between compartments is tailored around the epidemic data at hand, in both its qualitative and quantitative detail. 

The simplest, most general type of such models are SIR models, which consider only three compartments: Susceptible (those individuals in the population that can contract the illness); Infected (those who have been exposed to the illness, have themselves become ill and are acting as a source of epidemic spread); Recovered (those who have been sick and gained immunity, which prevents them from -- permanently or temporarily -- contracting the disease again). This is a simple concept which describes an epidemic system in an efficient way, via three variables $S(t)$, $I(t)$ and $R(t)$ (corresponding to each of the three compartments). Differential equations govern the interaction of these variables, specifying the likelihood (or the rate at which) susceptible people become infected, then recover, then eventually lose immunity and reenter the Susceptible compartment~\cite{zaman2007stability,ullah2013stability,kuniya2018stability}.
In a closed system (in which the birth and death rates are equal) the sum of
these compartments $N = S(t) + I(t) + R(t)$ remains constant in time. There are many different versions of SIR models that have been used in the literature to investigate a variety of epidemiological and behavioral aspects. In our work, we will consider the coupled dynamics of these compartments to be described by the following system of equations:

\begin{eqnarray*}
\frac{dS}{dt} &=&  \mu N - \frac{\beta S I}{N} - V - \mu S + \xi R \\
\frac{dI}{dt} &=&  \frac{\beta S I}{N} - \mu I - \gamma I  \\
\frac{dR}{dt} &=&  \gamma I + V - \mu R - \xi R\\
\end{eqnarray*}

\noindent where the variables and parameters are defined below.

Each coupled equation specifies the flow balance in one of the SIR compartments, respectively. The term $\mu N$ in the first equation accounts for new births. Our model operates under the assumption that the birth rate is proportional to the total population, and that babies are born healthy and with no virus-specific immunity. Corresponding death rates $-\mu S$, $-\mu I$ and $-\mu R$ were applied to all compartments. As mentioned before, this rate balance ensures that the population remains constant ($S+I+R=N$); in particular, it assumes that there are no illness-related fatalities (that is, all people recover). 

The transition rate $\frac{d(S/N)}{dt}$ at which the fraction of susceptible people in the population changes takes the form $\beta SI/N^2$.
Here, the constant parameter $\beta$ is the average number of contacts per person per time, multiplied by the probability of disease transmission via a contact between a susceptible individual, and an individual carrying the virus (and will be further discussed below). To obtain the transmission rate, this is multiplied by the fraction $SI/N^2$ representing the likelihood of an arbitrary contact to be between a susceptible and an infectious individual. 

As infected people progress through the symptoms and recover, their ability to spread the virus wanes. In our model, recovery  is assumed with bring perfect, but temporary immunity. Hence the transfer rate from the $I$ to the $R$ compartment is $\gamma I$, where $1/\gamma$ is the average recovery time. Since immunity is not permanent, the transfer rate from the Recovered to the Susceptible compartment is $\xi R$, where $1/\xi$ is the average immunity lifespan. In addition, the model was designed to also reflect the potential for individuals to obtain immunity without contracting the illness, via vaccination. This effect is introduced mathematically through a vaccination rate $V=V(t)$ between the Susceptible and Recovered compartments~\cite{chauhan2014stability}. Here, it was assumed that vaccination delivers immediate immunity; the vaccination-induced immunity is considered to be identical in duration with that conferred by actually having been infected. 

Our work on this variation of the SIR model will focus on investigating the effects of vaccination and health protection protocols, while keeping the other parameters fixed. We explore different options for implementing vaccination through $V(t)$. We consider different levels of health protective protocols by allowing $\beta$ to vary in a relatively large range. Since $\beta$ reflects the ``infectiousness'' of the virus, its value depends on the degree of adherence to social distancing measures in the community, with a lower $\beta$ when people are more careful, and a higher $\beta$ when people are more relaxed around social distancing measures, sanitizing, masking, etc~\cite{agrawal2016stability}. Unless otherwise specified, the values of the other system parameters (intrinsic to the epidemic dynamics) were held fixed to those in Table~\ref{parameter_values}.

\begin{table}[h!]
\begin{center}
\begin{tabular}{|l|l|l|l|}
\hline
{\bf Parameter name} & {\bf Significance} & {\bf Value or range} & {\bf Units}\\
\hline
$\beta$ & infectiousness & 0--1 & day$^{-1}$\\
\hline
$\nu$ & vaccination rate & 0--1 & day$^{-1}$\\
\hline
$\mu$ & birth/death rate & 0.0008 & day$^{-1}$\\
\hline
$1/\gamma$ & recovery time &  10--20 & days\\
\hline
$1/\xi$ & duration of immunity & 250 & days\\
\hline
\end{tabular}
\end{center}
\caption{\emph{{\bf Model parameters}, together with their significance in the model, units and the corresponding baseline value (for the fixed parameters) or the range (for the key parameters).}}
\label{parameter_values}
\end{table}

\section{Comparing different vaccination schemes}

A vaccination scheme used traditionally in SIR models is one that assumes inoculation at a constant rate. For a closed system in which $S+I+R = N$ is constant, this is the same as saying that a constant proportion $V(t) = \nu N$ leaves the susceptible compartment to join the Recovered compartment, due to vaccination. Since the birth rate in the model is also proportional to $N$, this can also be thought of as a fraction of the new population is constantly being vaccinated, as done in the case of some contagious diseases with planned vaccines (like measles or polio). However, this preventive approach is not always possible, and in reality vaccination needs to be implemented on a need-to basis, focusing on people who are at risk of becoming sick (i.e., the Susceptible population). In that case, the vaccination rate takes the form $V(t) = \nu S$, where $\nu$ refers instead to a fraction of the Susceptible compartment. 

In practice, both strategies are likely to be used at different times. For example, in the crisis of an ongoing outbreak, vaccination may have to aim for a fraction of the susceptible individuals (who are not currently sick or immune), while a longer-term plan, after the crisis was abated, can aim for vaccinating an equivalent fraction of the whole population. For simplicity, however, in this paper we investigate these two protocols separately and compare their efficiency. In future work, we plan to analyze further the combination of the two.

For both schemes, we computed the system's equilibria and used a standard linearization method to determine their local stability. We describe the results below.\\

\noindent \textbf{\emph{Equilibria and stability for constant vaccination scheme $V(t) = \mu N$.}} 

\noindent Starting with the equation
$I(\frac{\beta S}{N} - \mu - \gamma)=0$ to obtain the equilibrium component $I^*$, and continuing to solve for all components, one can easily compute two equilibria. For all parameter values, the system has an infection-free equilibrium ${\cal O}^*$ (in which the $I$ compartment is empty)
\begin{equation}
(S_0^*,I_0^*,R_0^*) = \left( \frac{(\mu+\xi-\nu)N}{\mu + \xi}, 0, \frac{\nu N}{\mu + \xi} \right)
\end{equation}

\noindent and an endemic equilibrium ${\cal E}^*$ (in which infection persists)
\begin{eqnarray}
S_E^* &=& \frac{(\mu + \gamma)N}{\beta}\\
I_E^* &=& \frac{N[\beta(\mu+\xi-\nu)-(\mu+\gamma)(\mu+ \xi)]}{\beta(\mu + \gamma +\xi)}\\
R_E^* &=& \frac{N[\beta(\gamma + \nu)-\gamma(\nu + \gamma)}{\beta(\mu + \gamma +\xi)}
\end{eqnarray}

Notice that, if $\nu > \mu + \xi$, neither equilibirum is biologically acceptable, since $S_0^*<0$, and $I_E^*<0$. In this case, the system will always approach a long-term state with $I^*=0$, either by converging to the infection-free state or by depleting the infected compartment in finite time while converging to the negative ``endemic'' state. Either way, this tells us that a high enough vaccination rate will always be guaranteed to clear infection. Below, we discuss the positivity and stability of the two equilibria when $\nu < \mu +\xi$.

By computing the Jacobian matrix at the infection-free equilibrium, one can obtain the eigenvalues directly as $\lambda_1 = -\mu$, $\lambda_2 = -(\mu+\xi)$, and 
$$\lambda_3 = \frac{\beta(\mu+\xi-\nu)-(\mu+\gamma)(\mu+\xi)}{\mu+\xi}$$

\noindent Based on this, we notice that the stability of the equilibrium $(S_0^*,I_0^*,R_0^*)$ changes from attracting node to saddle as the infectiousness parameter $\beta$ crosses past the critical value
\begin{equation}
\beta^* = \frac{(\mu+\gamma)(\mu+\xi)}{\mu+\gamma-\nu}
\label{transcritical_beta1}
\end{equation}

The linearization around the endemic equilibrium has one negative eigenvalue $\lambda_1 = -\mu$ and the other two eigenvalues given by the quadratic equation $A\lambda^2 + B\lambda + C=0$, where $A=\mu+\gamma+\xi$, $B=\beta(\mu+\xi-\nu)+\xi(\mu+\xi)$ and $C=(\mu+\gamma+\xi)(\beta(\mu+\xi-\nu)-(\mu+\gamma)(\mu+\xi)$. It follows that, when $\beta>\beta^*$, then $C<0$ and the endemic equilibrium is a saddle. When $\beta<\beta^*$, the equilibrium is a stable node if $B^2-4AC>0$ and a stable spiral if $B^2-4AC<0$. This condition defines two additional critical conditions for $\beta$, at
$$\beta_{\pm} = \frac{K \pm 2\sqrt{\gamma(\gamma+\mu+\xi)^3}}{\mu+\xi-\nu},$$

\noindent where $K = 2(\gamma+\mu)^2 + \xi(4\gamma+3\mu+\xi))$. 

Notice now that, if $\beta>\beta^*$, it follows that all components of the endemic equilibrium $(S_E^*,I_E^*,R_E^*)$ are positive. In addition, since $\beta^*<\beta_{-}$ for all parameter values, the following stability characterization results hold for the vaccination scheme we are considering:

\begin{thm}
If the infectiousness $\beta< \beta^*$, the infection-free equilibrium ${\cal O}^*$ is attracting, while the endemic equilibrium ${\cal E}^*$ is biologically implausible ($I_E^*<0$) and unstable (saddle). For $\beta>\beta^*$, ${\cal O}^*$ is unstable (saddle), and ${\cal E}^*$ is positive and attracting. The transition occurs via a transcritical bifurcation at $\beta=\beta^*$.
\end{thm}

\begin{thm}
For $\beta^*<\beta<\beta_{-}$ and $\beta>\beta_{+}$, the endemic equilibrium ${\cal E}^*$ is an attracting node. For $\beta_{-}<\beta<\beta_{+}$, then ${\cal E}^*$ is an attracting spiral, causing damped epidemic waves along the convergence process.
\end{thm}

\begin{figure}[h!]
\begin{center}
\includegraphics[width=0.48\textwidth]{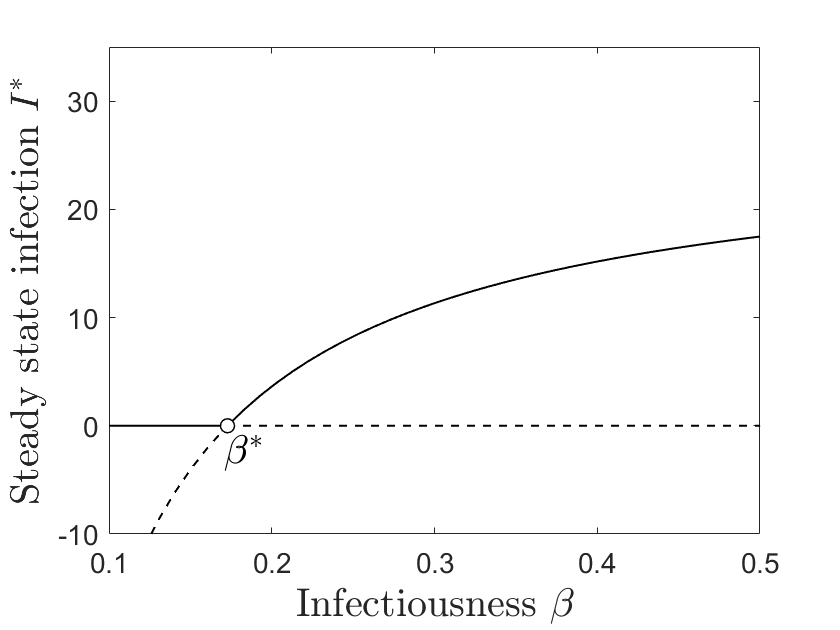}
\includegraphics[width=0.48\textwidth]{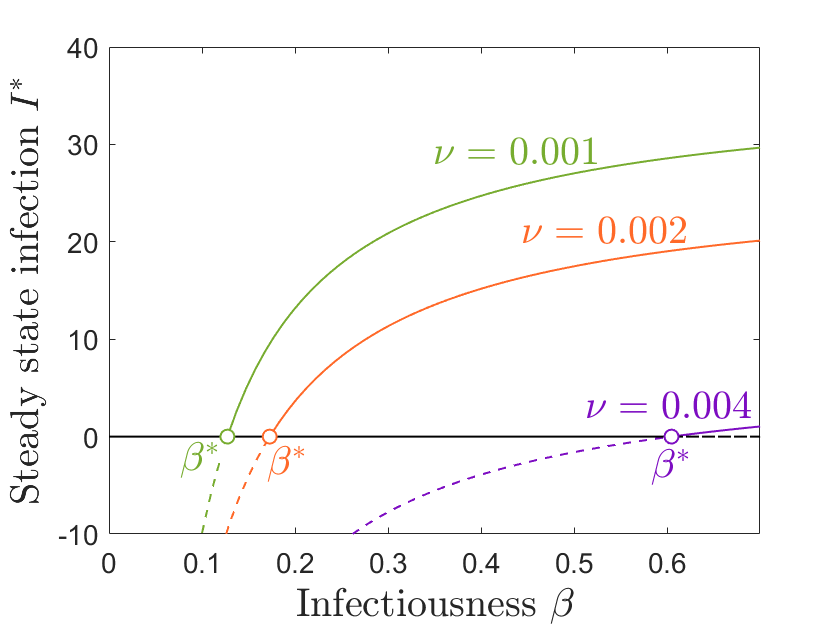}
\end{center}
\caption{\emph{{\bf Transcritical bifurcation} in terms of the infectiousness $\beta$. The vaccination rate was fixed to $\nu=0.002$. All other parameters were fixed to their table baseline values. {\bf Left.} Crossing of the infection-free and endemic equilibrium branches, with stability switching, at the transcritical bifurcation point $\beta^* \sim 0.17$. {\bf Right.} Effect on the equilibrium curves and on the position of the transcritical bifurcation when the vaccination rate $\nu$ is varied from $\nu = 0.001$ (green curves, for which $\beta^* \sim 0.12$) to $\nu=0.002$ (orange curves, for which $\beta^* \sim 0.17$, as in the left panel) to $\nu=0.004$ (purple curves, for which $\beta^* \sim 0.6$).}}
\label{transcritical_scheme1}
\end{figure}

Figure~\ref{transcritical_scheme1} illustrates the evolution of the equilibria ${\cal O}^*$ and ${\cal E}^*$ as $\beta$ increases and crosses the transcritical bifurcation at $\beta ^*$. It also shows how these curves and the position of the bifurcation are affected when the vaccination rate is increased twice by a factor of two each time. As one would expect intuitively, as well as from equation~\ref{transcritical_beta1}, higher $\nu$ values lead to higher transcitical values $\beta^*$ and higher endemic infection levels for $\beta>\beta^*$. This means that, when vaccination is increased, convergence to the infection-free state will be ensured up to higher infectiousess values $\beta$ (which may correspond to slightly more relaxed health protocols). In other words, there is a tight trade-off between the population's compliance with health versus vaccination protocols. Being stricter with one may allow the community to be more relaxed with the other, and still be able to control or even clear the infection in the long term.


\vspace{5mm}
\noindent \textbf{\emph{Equilibria and stability for scaled vaccination scheme $V(t) = \nu S$.}} We will work out a similar analysis for this case. Notation-wise, we will use a hat here to distinguish the equilibria and transitions from the corresponding ones in the first vaccination scheme.

\vspace{2mm}
\noindent As before, the system has an infection-free equilibrium $\hat{\cal O}^*$ with components
\begin{equation}
(\hat{S}_0^*,\hat{I}_0^*,\hat{R}_0^*) = \left( \frac{(\mu+\xi)N}{\mu + \xi+\nu}, 0, \frac{\nu N}{\mu + \xi + \nu} \right)
\end{equation}

\noindent and an endemic equilibrium $\hat{\cal E}^*$
\begin{eqnarray}
\hat{S}_E^* &=& \frac{(\mu + \gamma)N}{\beta}\\
\hat{I}_E^* &=& \frac{N[\beta(\mu+\xi)-(\mu+\gamma)(\mu+ \xi+\nu)]}{\beta(\mu + \gamma +\xi)}\\
\hat{R}_E^* &=& \frac{N[\beta \gamma +(\nu-\gamma)(\mu + \gamma)}{\beta(\mu + \gamma +\xi)}
\end{eqnarray}

\noindent The eigenvalues of the Jacobian matrix at $\hat{\cal O}^*$ are $\lambda_1 = -\mu$, $\lambda_2 = -(\mu+\xi+\nu)$ and 
$$\lambda_3 = \frac{\beta(\mu+\xi)-(\mu+\gamma)(\mu+\xi+\nu)}{\mu+\xi+\nu}$$

\noindent As before, the stability of the equilibrium $(\hat{S}_0^*,\hat{I}_0^*,\hat{R}_0^*)$ changes from attracting node to saddle as the endemic equilibrium $\hat{\cal E}^*$ becomes biologically relevant (all positive components) -- that is, when the infectiousness parameter $\beta$ crosses past the critical value
$$\hat{\beta}^* = \frac{(\mu+\gamma)(\mu+\xi+\nu)}{\mu+\gamma}$$

\noindent At this transcritical bifurcation, the stability of the endemic equilibrium $\hat{\cal E}^*$ also changes from unstable/saddle (for $\beta<\beta^*$) to stable (for $\beta>\beta^*$). Further tracking the transition between stable node and stable spiral, one can compute 
$$\hat{\beta}_{\pm} = \frac{\hat{K} \pm 2\sqrt{(\gamma-\nu)(\gamma+\mu+\xi)^3}}{\mu+\xi},$$

\noindent where $\hat{K} = 2(\gamma+\mu)^2 + \xi(4\gamma+3\mu+\xi-\nu))$. As before, one obtains a stability characterization result:

\begin{thm}
If the infectiousness $\beta< \hat{\beta}^*$, the infection-free equilibrium $\hat{\cal O}^*$ is attracting, while the endemic equilibrium $\hat{\cal E}^*$ is biologically implausible ($\hat{I}_E^*<0$) and unstable (saddle). For $\beta>\hat{\beta}^*$, $\hat{\cal O}^*$ is unstable (saddle), and $\hat{\cal E}^*$ is positive and attracting. The transition occurs via a transcritical bifurcation at $\beta=\hat{\beta}^*$.
\end{thm}

\begin{thm}
For $\hat{\beta}^*<\beta<\hat{\beta}_{-}$ and $\beta>\hat{\beta}_{+}$, the equilibrium $\hat{\cal E}^*$ is an attracting node. For $\hat{\beta}_{-}<\beta<\hat{\beta}_{+}$, then $\hat{\cal E}^*$ is an attracting spiral, causing damped epidemic waves along the convergence process.
\end{thm}

\begin{figure}[h!]
\begin{center}
\includegraphics[width=0.48\textwidth]{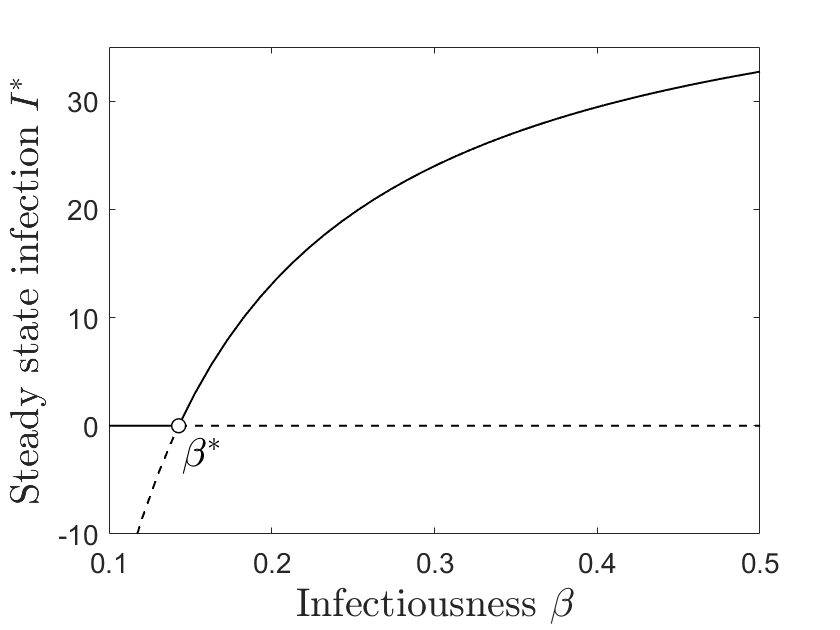}
\includegraphics[width=0.48\textwidth]{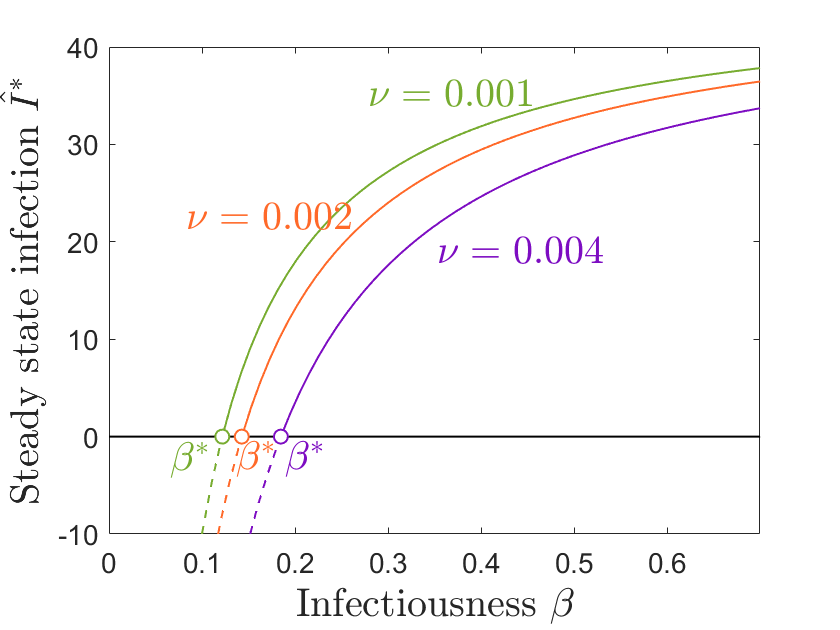}
\end{center}
\caption{\emph{{\bf Transcritical bifurcation} in terms of the infectiousness $\beta$. The vaccination rate was fixed to $\nu=0.002$. All other parameters were fixed to their table baseline values. {\bf Left.} Crossing of the infection-free and endemic equilibrium branches, with stability switching, at the transcritical bifurcation point $\beta^* \sim 0.14$. {\bf Right.} Effect on the equilibrium curves and on the position of the transcritical bifurcation when the vaccination rate $\nu$ is varied from $\nu = 0.001$ (green curves, for which $\beta^* \sim 0.12$) to $\nu=0.002$ (orange curves, for which $\beta^* \sim 0.14$, as in the left panel) to $\nu=0.004$ (purple curves, for which $\beta^* \sim 0.18$).}}
\label{transcritical_scheme2}
\end{figure}

Figure~\ref{transcritical_scheme2} illustrates in this case the evolution of the equilibria ${\cal O}^*$ and ${\cal E}^*$ as $\beta$ increases and crosses the transcritical bifurcation at $\beta ^*$. We also show how these curves change when the vaccination rate $\nu$ takes the same three values as in the first scheme. Higher $\nu$ values are proving again to be more efficient at controlling and eliminating infection, even under higthtened infectiousness conditions. However, while the general effect if similar between the two vaccination schemes, there are also significant differences, which we discuss below.


\vspace{5mm}
\noindent \textbf{\emph{Comparison between the two vaccination schemes.}}

\noindent We want to compare the efficiency of the two vaccination schemes operating with the same value $\nu$ (the vaccinated fraction of $N$ and $S$, respectively). We will focus on comparing the three most important features of these dynamics, which can be viewed to encompass the long-term epidemic signature of the system: (1) the position of the transcritical bifurcation with respect to $\beta$ and (2) the persistent infection corresponding to the stable endemic equilibrium, for $\beta$ beyond this bifurcation.
As mentioned before, the position of the bifurcation is of practical importance since it defines limitations for people's behavior with respect to health and social distancing measures (the relaxation of which may increase $\beta$ to the endemic range). The level of endemic infection is also crucial since it defines the persistent prevalence of infection in the community. In addition, knowledge on whether or not, and to what extent, the endemic infection performs wide, long-term oscillations around the steady state, could be of major importance when preparing the health care system for period of infection resurgence.

A direct calculation shows that, for the same value of $\nu$, we have both that $\beta^*(\nu)<\hat{\beta}^*(\nu)$ and that $I^*(\nu) > \hat{I}^*(\nu)$ for all parameter values. This supports with evidence the intuition that constant vaccination scheme outperforms the scaled vaccination scheme in both ``delaying'' the  endemic state (in terms of providing a larger allowance for misgivings in health protocols) and minimizing the size of the endemic infection. Next, we plan to study how the picture changes if one considers feedback adaptation health protocols by taking the value of $\beta$ to depends on the current state of infection. 

\begin{figure}[h!]
\begin{center}
\includegraphics[width=0.48\textwidth]{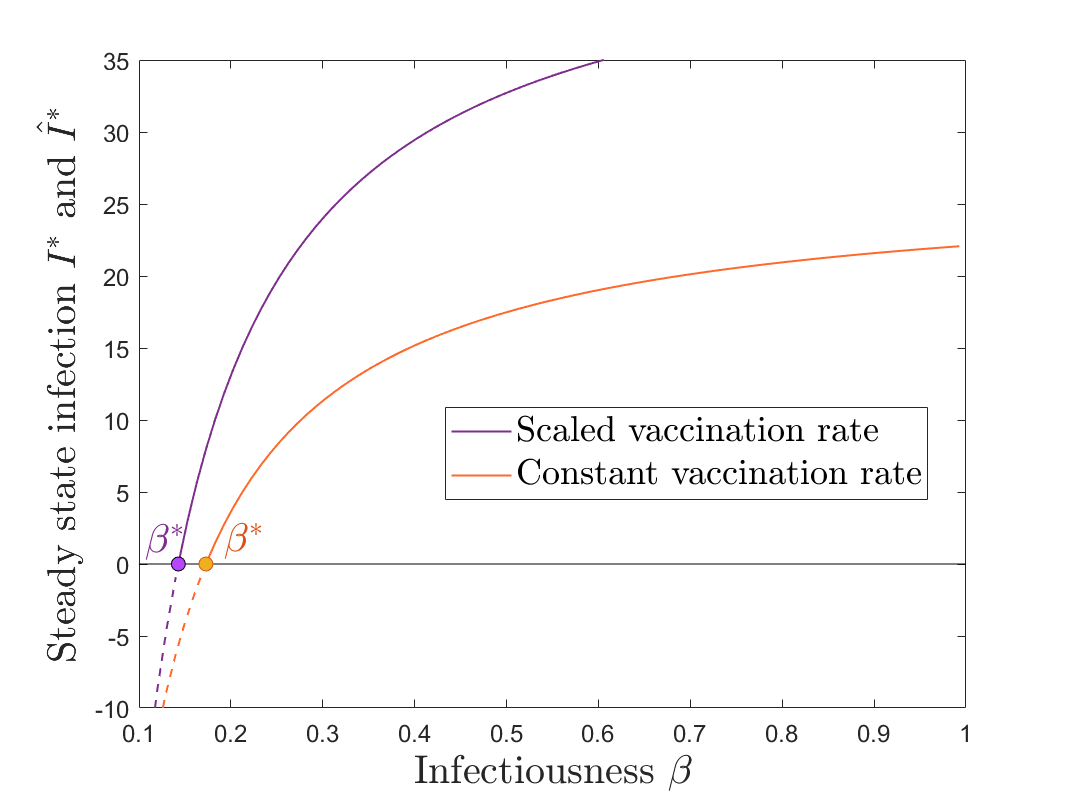}
\includegraphics[width=0.48\textwidth]{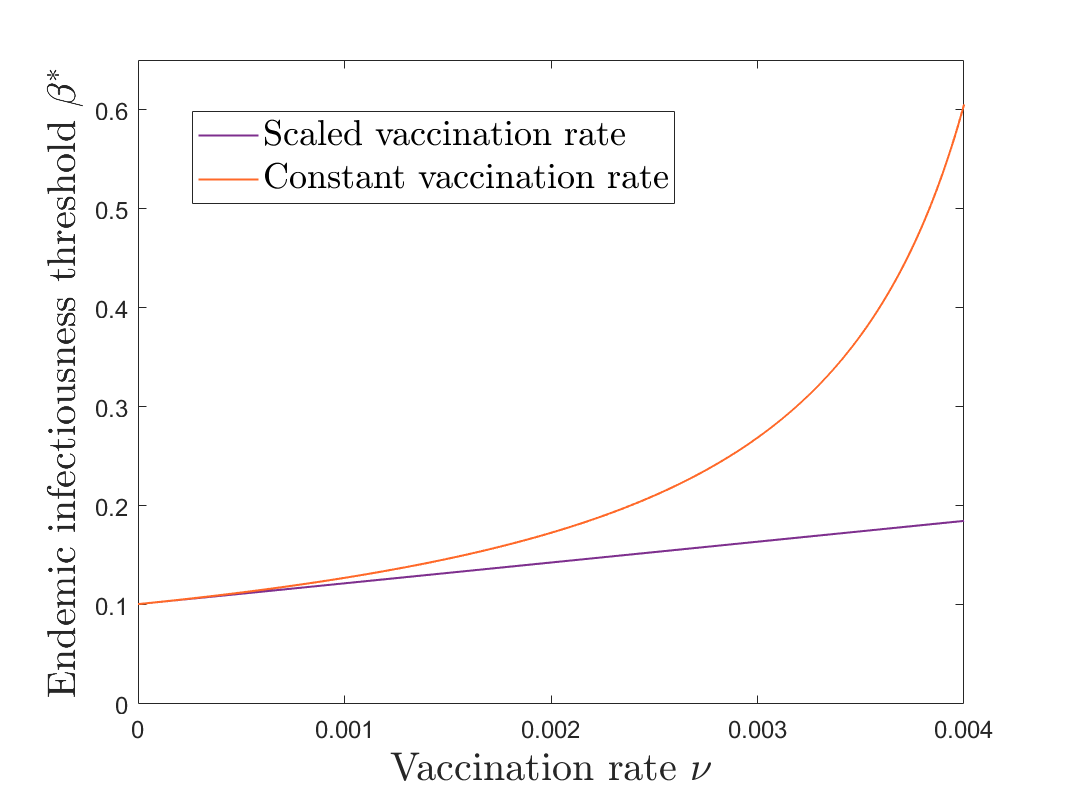}
\end{center}
\caption{\emph{{\bf Comparison between vaccination schemes.} {\bf Left.} The evolution of the endemic equilibrium is shown in terms of $\nu$, for fixed parameters $\gamma=0.1$, $\mu = 0.0008$, $\xi=0.004$. For the same value of $\nu$, the constant rate scheme always delivers a higher value for the transcritical bifurcation $\hat{\beta}^*(\nu)>\beta^*(\nu)$, and lower endemic infection $\hat{I}^*(\nu)<I^*(\nu)$. {\bf Right.} Comparison between $\beta*(\nu)$ (purple curve) and $\hat{\beta}^*(\nu)$ (orange curve), for the same fixed parameters.}}
\label{comparison}
\end{figure}


\section{Population behavior around social distancing protocols} 

The previous section gave us an idea of the importance of the interplay between vaccination and health protective protocols when trying to end or mitigate an epidemic outbreak. In this section, we will further illustrate this interplay under two extreme scenarios of behavior during an epidemic outbreak. We analyzed these two scenarios in conjunction with both vaccination schemes discussed in the previous section.

In the first scenario, the population adapts their behavior to increase their adherence to health protocols and social distancing rules (effectively decreasing $\beta$) with increasing infection in the community. This form of negative feedback is rooted in the current reality of the population having almost instant access to incidence data in the community, and being able to use it to make informed decisions on mobility and social behavior. The effect may be augmented by official mandates, sometimes enforced when dealing with high infection during a particularly concerning outbreak. 

In the second scenario, the population decreases their adherence to health protocols with increasing infection. While this may seem improbably and go against common sense, it may occur temporarily, when there exists an epidemic-independent, strong trigger for relaxation in protocols, that transcends the natural tendency to apply the negative feedback described earlier. A prominent and planned social event in the community, or simply the winter holidays, may lead to people letting their guard down (and effectively increasing $\beta$) as cases are increasing.

To mathematically implement both the negative feedback and the reinforcing feedback scenarios in one comprehensive, realistic, but simple enough way -- we set the model infectiousness $\beta$ to adapt with infection, as a nonlinear, sigmoidal-shaped function:
\begin{equation}
\beta(I) = \beta_0 \mp \frac{\beta_{\max}}{1+\exp[-k_{\max}(I-I_{\max})]}
\end{equation}

\begin{figure}[h!]
\begin{center}
\includegraphics[width=0.48\textwidth]{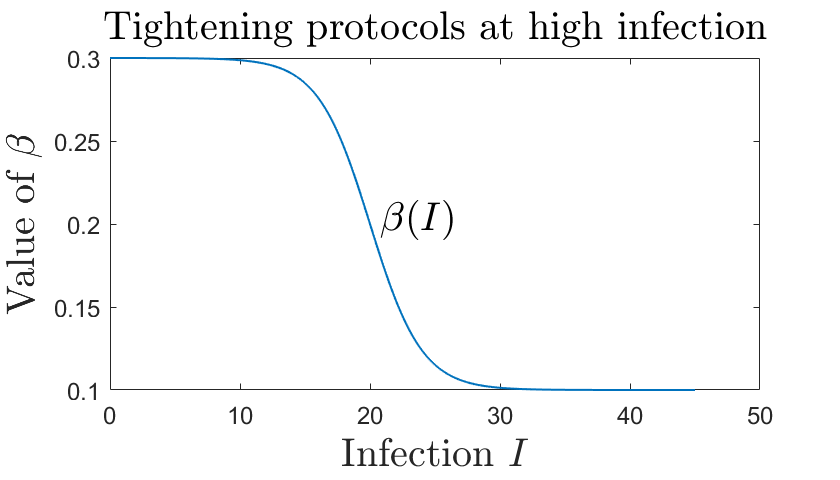}
\includegraphics[width=0.48\textwidth]{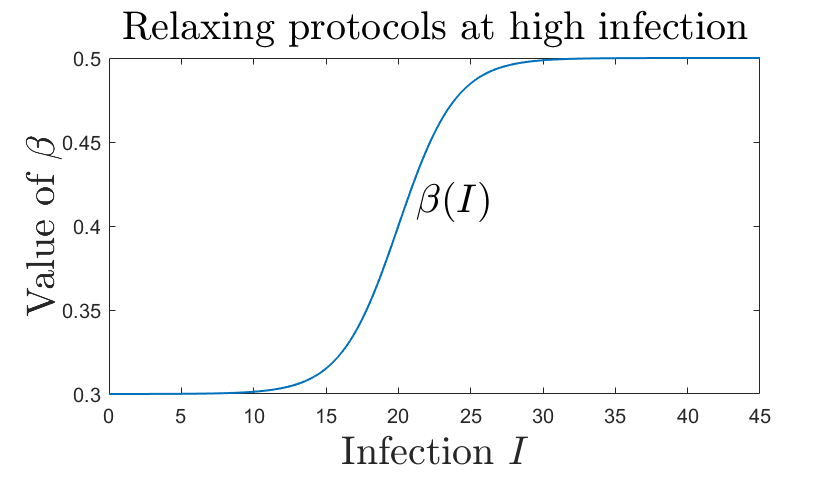}
\end{center}
\caption{\emph{{\bf Dependence of $\beta$ on the current infection state.} {\bf Left.} As infection increases, the infectiousness is decreasing: slowly at first, then more sensitively with $I$, then tappers off towards a minimum values (in this case, $\beta_0-\beta_{\max} = 0.1$. {\bf Right.} As infection increases, the infectiousness is increasing slowly, then faster, then saturates to an asymptotic value (in this case, $\beta_0+\beta_{\max} = 0.5$.}}
\label{sigmoidals}
\end{figure}

\noindent Here, $\beta_0$ is the baseline value of the infectiousness in absence of adaptation; $\beta_{\max}<\beta_0$ is the magnitude of the maximum adjustment possible in response to increasing infection (and is applied subtractively or additively, depending on the scenario). $I_{\max}$ is the position of the inflection point of the function, that is the infection level where $\beta$ shows the highest sensitivity to $I$ (a small change in infection produces a dramatic adaptation in $\beta$. Finally, $k_{\max}$ is the slope of the function $\beta(I_{\max})$. For illustration purposes, in our simulations we used the following values for these modulation parameters: $\beta_0=0.3$, $\beta_{\max} = 0.2$, $I_{\max}=0.02N = 20$ and $k_{\max}=0.5$. The sigmoidal graphs for both decreasing and increasing $\beta(I)$ scenarios are shown in Figure~\ref{sigmoidals}.

Notice that finding equilibria of the system with one of these two adaptations can still be approached by first solving the algebraic  equation: 

$$\frac{\beta SI}{N} - \mu I -\gamma I = 0$$

\noindent which leads to an infection-free equilibrium, and whichever other equilibria are generated from solving
\begin{equation*}
(\mu+\xi-\nu) -(\mu+\gamma+\xi)I = \frac{(\mu+\xi)(\mu+\gamma)}{\beta(I)}
\end{equation*}

Notice that the left side is a line with negative slope with respect to $I$. Hence, if $\beta(I)$ decreases with $I$ (our first scenario), the equation has only one solution, and the system has only one possible endemic equilibrium. However, is $\beta(I)$ increases with $I$ (our second scenario), the system can have in principle any number of endemic equilibria. We study  these aspects in more detail below, through numerical simulations, for both vaccination schemes discussed in the previous section.


\subsection{Adaptable behavior with scaled vaccination scheme}
\quad

\vspace{2mm}
\noindent We set to investigate the importance of vaccination in conjunction with each of the two adaptations (decreasing and increasing $\beta(I)$), by observing the change in the system's long-term state as the vaccination rate $\nu$ is increased.\\ 

\vspace{2mm}
\noindent \textbf{\emph{ Increasing vaccination when the infectiousness $\beta$ decreases with case incidence.}} 

\noindent Our numerical simulations (performed with the Matcont software~\cite{}), suggest that the system always has two equilibria, one infection-free ${\cal O}$ and one endemic ${\cal E}$. Details of their behavior are shown in Figure~\ref{decreasing_beta}, for two different recovery rates ($\gamma=20$, panels b, and $\gamma=10$, panel c respectively). Notice that, in both cases, high enough vaccination rates eliminate infection , since ${\cal O}$ is the only biologically plausible equilibrium (the endemic state has $I_E^*<0$), and is also locally stable. As the vaccination rate is decreased, the system undergoes a transcritical bifurcation at the critical rate $\nu^*=0.025$, where ${\cal O}$ loses stability and  ${\cal E}$ becomes positive and locally attracting, thus representing for $\nu<\nu^*$ the only possible long-term prognosis for the outbreak. Notice that the lower the vaccination rate, the poorer the outcome (i.e., the higher the persistent infection $I_E^*$ in the community).

\begin{figure}[h!]
\begin{center}
\includegraphics[width=0.48\textwidth]{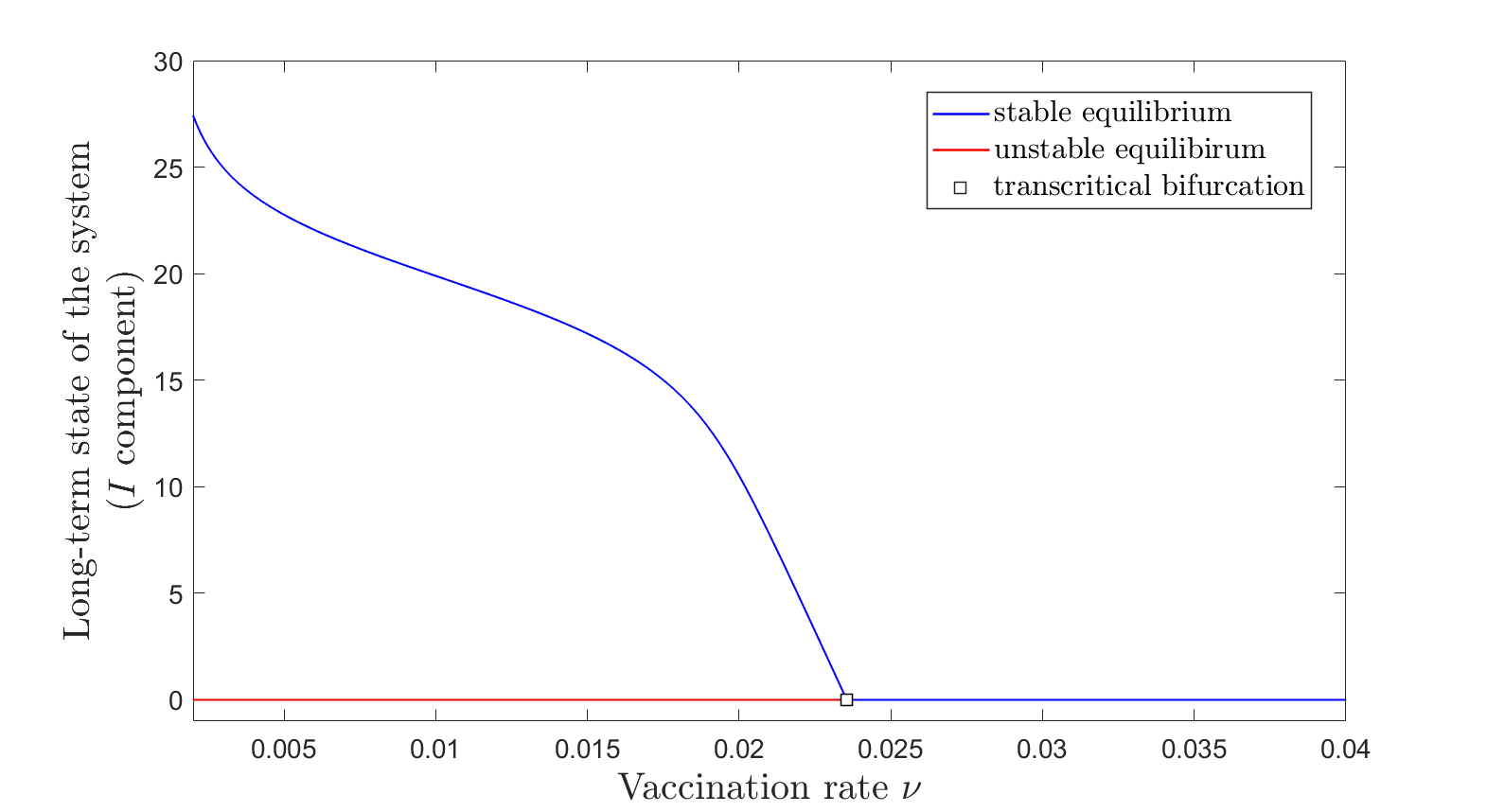}
\includegraphics[width=0.48\textwidth]{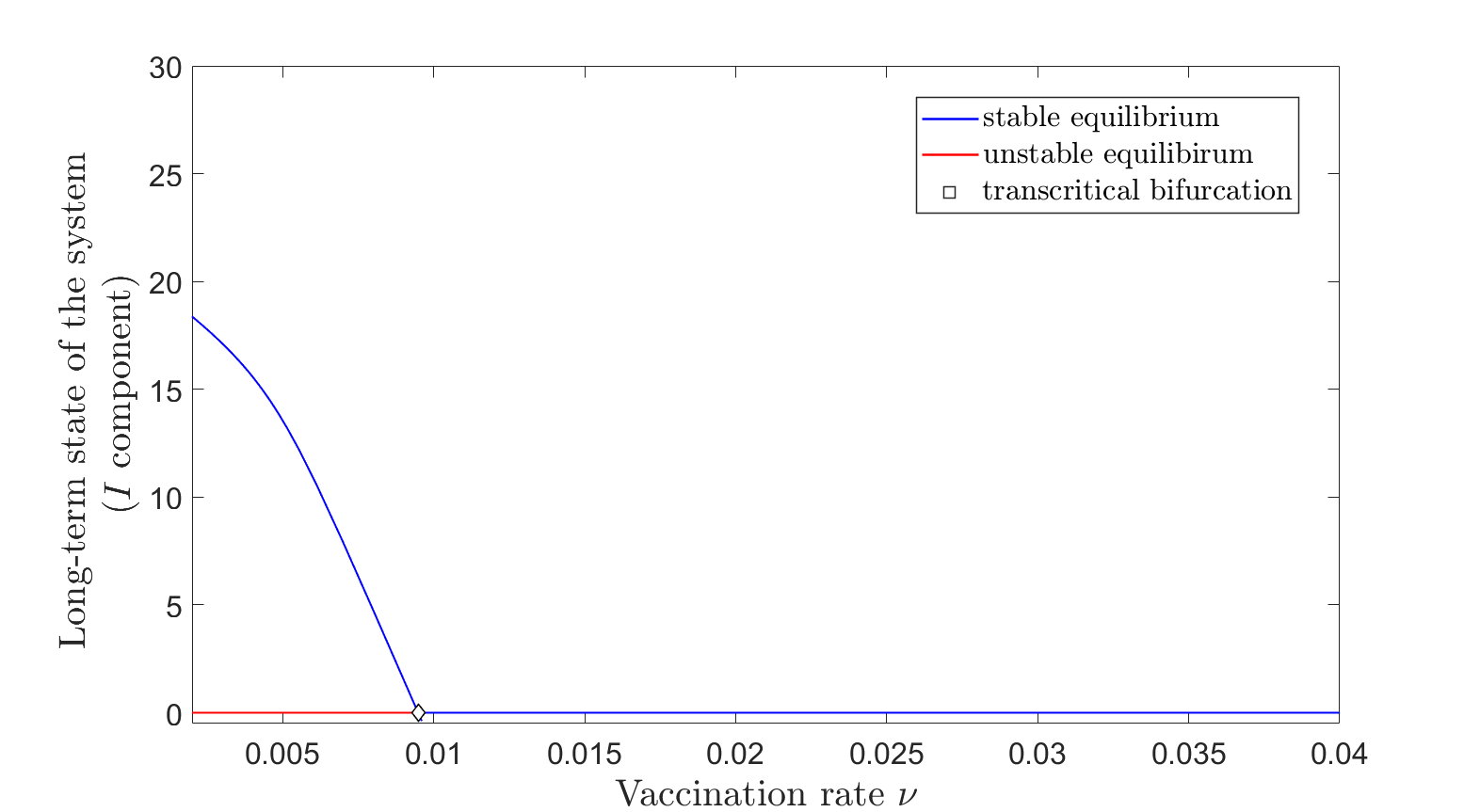}
\caption{\emph{{\bf Equilibria and bifurcations with respect to $\nu$, in the case of decreasing $\beta(I)$.} The $I$ component of the equilibirium branches is plotted in blue (if the equilibrium is stable) and in red (if the equilibrium is unstable). A transcritical bifurcation occurs at $\nu^*$, marked on the graphs as a white square. {\bf Left.} If the recovery period is $1/\gamma = 20$ days, $\nu^* \sim 0.025$. {\bf Right.} If the recovery period is $1/\gamma = 20$ days, $\nu^* \sim 0.01$.}}
\label{decreasing_beta}
\end{center}
\end{figure}

This is not an unexpected conclusion. The model predicts in this case that a vaccination rate as low as $\nu=0.01$ (equivalent to $\nu N =10$ individuals vaccinated daily) is sufficient to insure clearance from an epidemic illness which takes $1/\gamma =10$ days for recovery. If the recovery time is $1/\gamma = 20$ days, the critical vaccination rate that guarantees clearance increases to $\nu = 0.025$ (that is, $\nu N = 25$ individuals vaccinated per day). A suboptimal vaccination rate will lead in each case to endemic infection: the lower the rate, the more prevalent the cases.\\

\noindent \textbf{\emph{ Increasing vaccination when the infectiousness $\beta$ increases with case incidence.}} 

\noindent Our numerical simulations support the original idea that an increasing $\beta(I)$ defines a much more complex mathematical context. Not only the equilibria and transitions are richer, but the picture if also more dependent on other system parameters. For example, we obtained two qualitatively different outlines for shorter and longer recovery rates. In reality, this implies that these is a need in this case to obtain finer information about the system before making predictions and deciding on a vaccination protocol.

Indeed, our simulations for the system with a recovery rate of 20 days reveal the presence of a bistability window and hysteresis, as shown in Figure~\ref{increasing_beta}a. As before, a high enough vaccination rate insures convergence to an infection-free state, since ${\cal O}$ is the only biological equilibrium, and it is asymptotically stable. However, if the vaccination rate is lowered past $\nu^* \sim 0.03$, the system undergoes a saddle-node bifurcation (white diamond), which gives birth to a pair of equilibria: one is a stable node (which we will call ${\cal E}^*_1$), and the other one is unstable (saddle). The endemic state ${\cal E}^*_1$ now coexists with the infection-free state ${\cal O}$, with convergence to one or the other being determined by the initial conditions. This situation persists as vaccination decreases, until ${\cal O}^*$ loses stability through a transcritical bifurcation (white square) at $\nu^* \sim 0.025$. The role of stable equilibrium is picked up from ${\cal O}^*$ by a second endemic state ${\cal E}^*_2$, which now coexists with ${\cal E}^*_1$. The bistability window ends at $\nu^* \sim 0.02$, where ${\cal E}^*_2$ collides with the saddle equilibrum at a second saddle-node bifurcation (white diamond) and disappears. For vaccination rates $\nu<0.02$, the prognosis of the system is determined by the only remaining stable state, ${\cal E}^*_1$. 

\begin{figure}[h!]
\begin{center}
\includegraphics[width=0.48\textwidth]{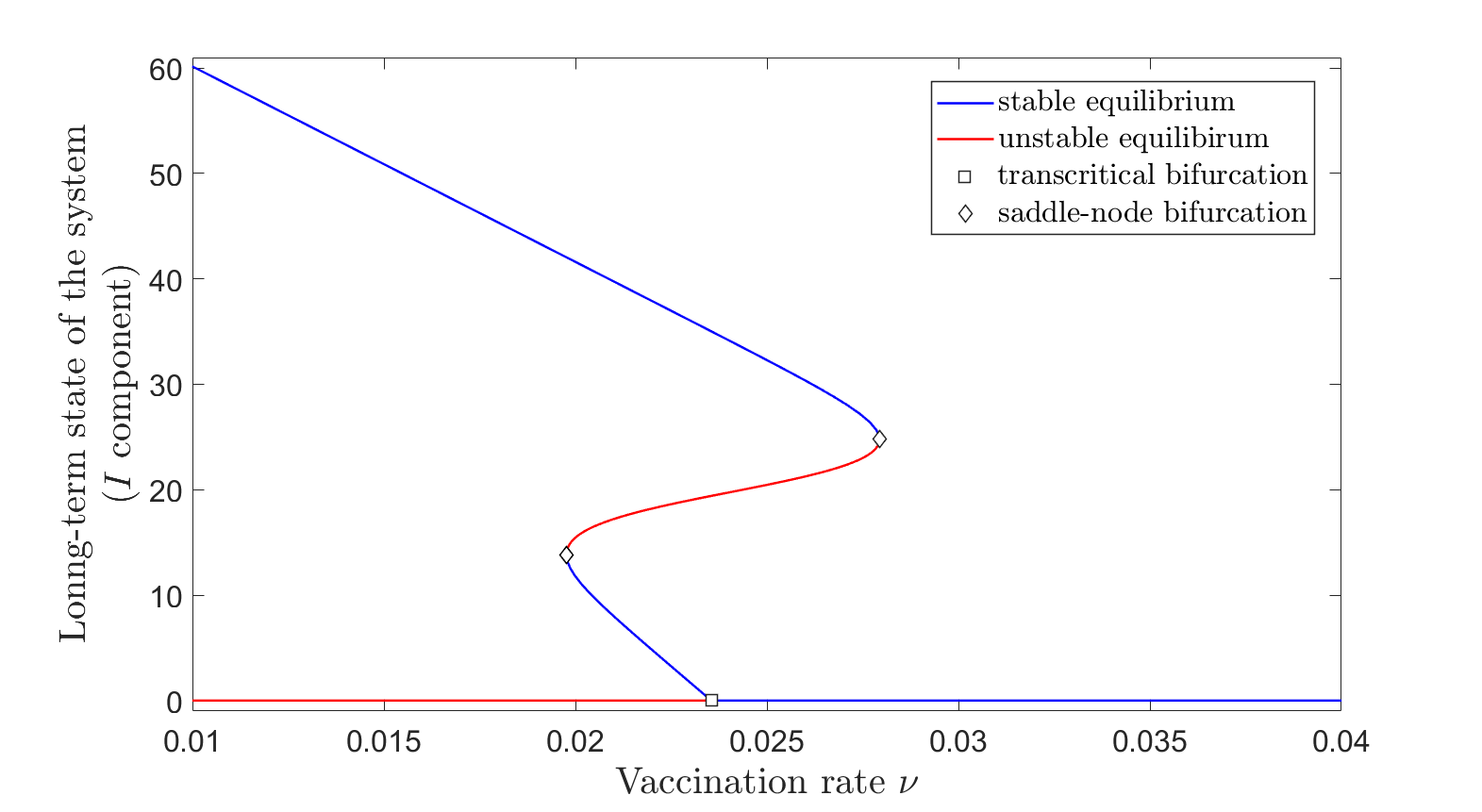}
\includegraphics[width=0.48\textwidth]{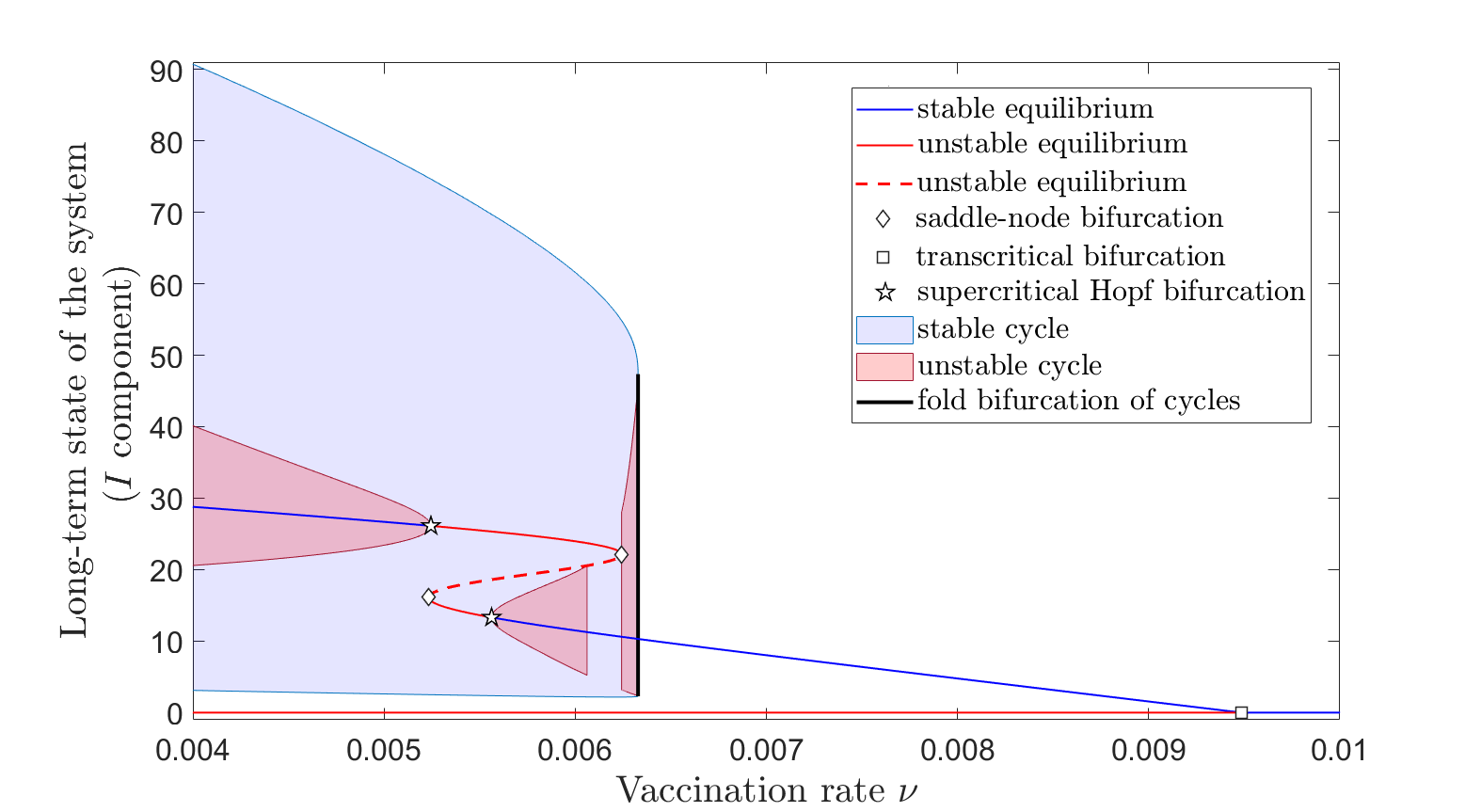}
\caption{\emph{{\bf Equilibria and bifurcations with respect to $\nu$, in the case of decreasing $\beta(I)$.} The $I$ component of the equilibirium branches is plotted in blue (if the equilibrium is stable) and in red (if the equilibrium is unstable). The formation and evolution of stable and unstable cycles are also shown, as their projection in the $I$ component. For a given $\nu$, the representation of the cycle is then a vertical segment, with its top and bottom (i.e., its min and max infection levels) evolving with $\nu$. Then the evolution of the cycle itself is shown as a shaded region between these bounding curves. Bifurcations are represented as follows: transcritical (white square); saddle node (white diamond); supercritical or subcritical Hopf (white star); fold, or saddle node of cycles (black bar). {\bf Left.} Bifurcation diagram for recovery period $1/\gamma = 20$ days. As vaccination is decreased from $\nu>0.03$ to $\nu<0.01$, the system transitions from convergence to an infection-free state, to two-state bistability, and eventually to a high-infection endemic state. {\bf Right.} Bifurcation diagram for recovery period $1/\gamma = 20$ days. As vaccination is decreased from $\nu=0.01$ to $\nu=0$, the system transitions, via a complex bifurcation sequence, from convergence to an infection-free state, to a low infection endemic state, then eventually to a bistability window in which an endemic equilibrium coexists with a stable cycle (epidemic waves).}}
\label{increasing_beta}
\end{center}
\end{figure}

Altogether, this shows how the prognosis of the model degrades as vaccination rates are decreased. High vaccination rates guarantee infection clearance. Lowering the rate introduces the risk of endemic infection, risk increased if the community is in a high-infection state to start with. Eventually, the community is doomed to have endemic infection, with lower $\nu$ leading to higher levels of persistent infection.

We observed a very different picture of attractors and transitions between them when the recovery time was decreased to $1/\gamma = 10$ days. A shorter time is advantageous, because recovered individuals are no longer spreaders, hence this situation leads to decreasing potential exposures by comparison with the $1/\gamma=20$ day recovery. Our simulation supports this advantage, revealing the infection- free ${\cal O}^*$ to be the only attractor for vaccination rates as low as $\nu = 0.01$. Even as the vaccination crosses below the transcritical bifurcation level $\nu^* \sim 0.01$ (white square), the system will converge to an endemic equilibrium with relatively low infection (we will call this ${\cal E}^*_1$). This equilibrium survives down to $\nu^* \sim 0.0055$, where the stability is destroyed by a collision with a saddle cycle, via a supercritical Hopf bifurcation (white star). After two saddle-node bifurcation in which the stability is further degraded and then regained, the stability of the equilibrium is restored through another supercritical Hopf bifurcation, at $\nu^* \sim 0.005$. However, the evolution of the equilibrium ${\cal E}^*_1$ does not tell the entire dynamics story. Starting at $\nu^* \sim 0.0063$, ${\cal E}^*_1$ coexists with a stable cycle, formed via a fold bifurcation, or saddle node for cycles (black vertical bar), and which survives all the way to $\nu=0$. 

In this scenario, vaccination as low as the transcritical point clears the infection completely. Vaccination as low as the fold point leads to a relatively low endemic infection level (up to 10 individuals out of a community of 1000). When decreasing vaccination below the fold point, perpetual epidemic waves become possible, as the only other alternative to endemic infection, with amplitude depending on $\nu$. In the intervals of bistability where the endemic state and the waves are both possible, convergence to one or the other is governed by the initial state of the system.\\

\noindent Before we move on to analyze the competing vaccination scheme, we would like to draw a few conclusions. First, one would certainly expect that the epidemic parameters may have a significant impact on the prognosis of the system and on the effects of vaccination. We illustrated this idea by comparing bifurcation diagrams between two different recovery times (10 and 20 days), pointing out the quantitative and qualitative differences between them. Understanding these differences in a real outbreak would provide priceless information on tuning the vaccination efforts according to the situation.

Second, we found remarkable differences between the two behavioral scenarios for $\beta(I)$. As expected from our analytical results, a community compliant with health protocols during high infection lead to a simple picture with only two equilibria, swapping stability at a transcritical bifurcation: the infection0free one prevails at higher vaccination rates, and the endemic one prevails at lower vaccination rates. In this case, even a poor vaccination protocol results in relatively low endemic infection (20-50 individuals in a community of N=1000, if no vaccination is implemented). Supporting both our previous work and others', this result supports the importance of adherence to health protocols, and their potential success at mitigating epidemics, even in vaccination non-compliant communities. In contrast, our model prediction for a population that behaves against health protocols includes higher persistent infection (up to 80-90 individuals in N=1000 when no vaccines are implemented, and 20-30 infections for vaccination rates range between 10-30 people per day, depending on the recovery period). It also includes the potential for periodic epidemic waves, with approximately annual occurrence, and peaking at as highly as 60-90 infections, the lower the vaccination rates, the higher the peaks).

Third, we need to keep in mind that these bifurcation diagrams should not be regarded only as a static predictive instrument (predicting the outcome of vaccinating at a particular rate), but also as a tool to create dynamics, adaptive strategies. For example, suppose the system in 
Figure~\ref{increasing_beta}a operates within the bistability window, for a  $0.02<\nu<0.023$, and is converging to the high endemic equilibrium ${\cal E}^*_1$. Simply applying a short boost in vaccination rate to a level higher than the saddle node value $\nu^*=0.03$ would give the system convergence time towards the infection-free equilibrium ${\cal O}^*$. This adjustment may be enough to move the system's state from the attraction basin of ${\cal E}^*_1$ into the attraction basin of ${\cal O}$, once returned to the original vaccination rate. This example, as well as many other similar ones, support the idea that even temporary increases in vaccination, applied strategically, can be very effective.

\subsection{Adaptable behavior with constant vaccination scheme}
\quad

\noindent \textbf{\emph{ Increasing vaccination when the infectiousness $\beta$ decreases with case incidence.}} 

\noindent The dynamics observed were qualitatively very similar to those for scaled vaccinations, with only two equilibria (infection-free ${\cal O}^*$ and endemic ${\cal E}^*$). One analytical difference consists of the fact that high vaccination rates lead to negative steady state values in the S and I compartments, which are not biologically plausible. However, from the perspective of the application, this means that the infection is eliminated in finite time, and the model ceases to work after that (since there are no people left to vaccinate). As before, a detailed behavior of the equilibria is shown in Figure~\ref{decreasing_beta}, for two different lengths of recovery ($\gamma^{-1}=20$ days, panels b, and $\gamma^{-1}=10$ days, panel c respectively). 

\begin{figure}[h!]
\begin{center}
\includegraphics[width=0.48\textwidth]{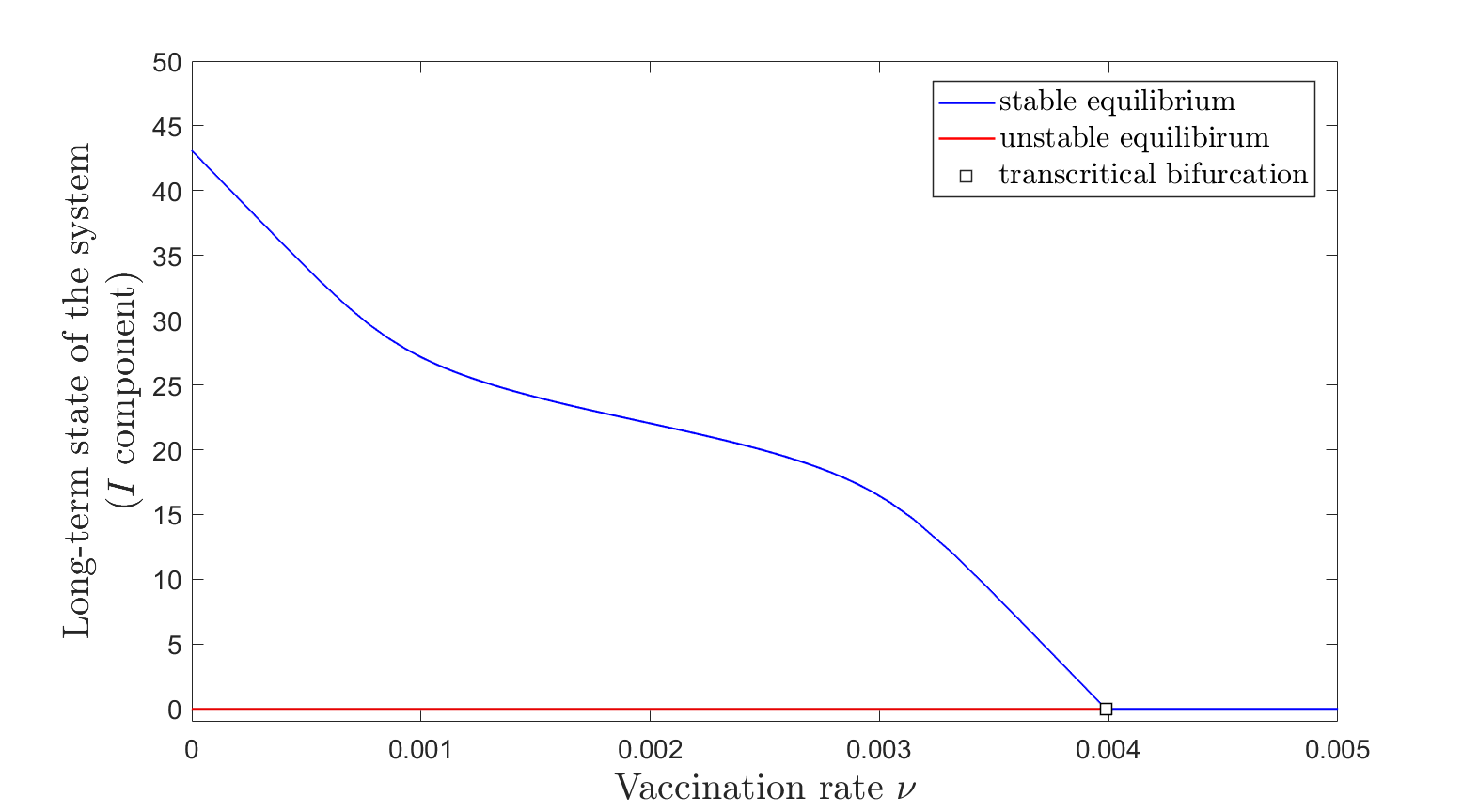}
\includegraphics[width=0.48\textwidth]{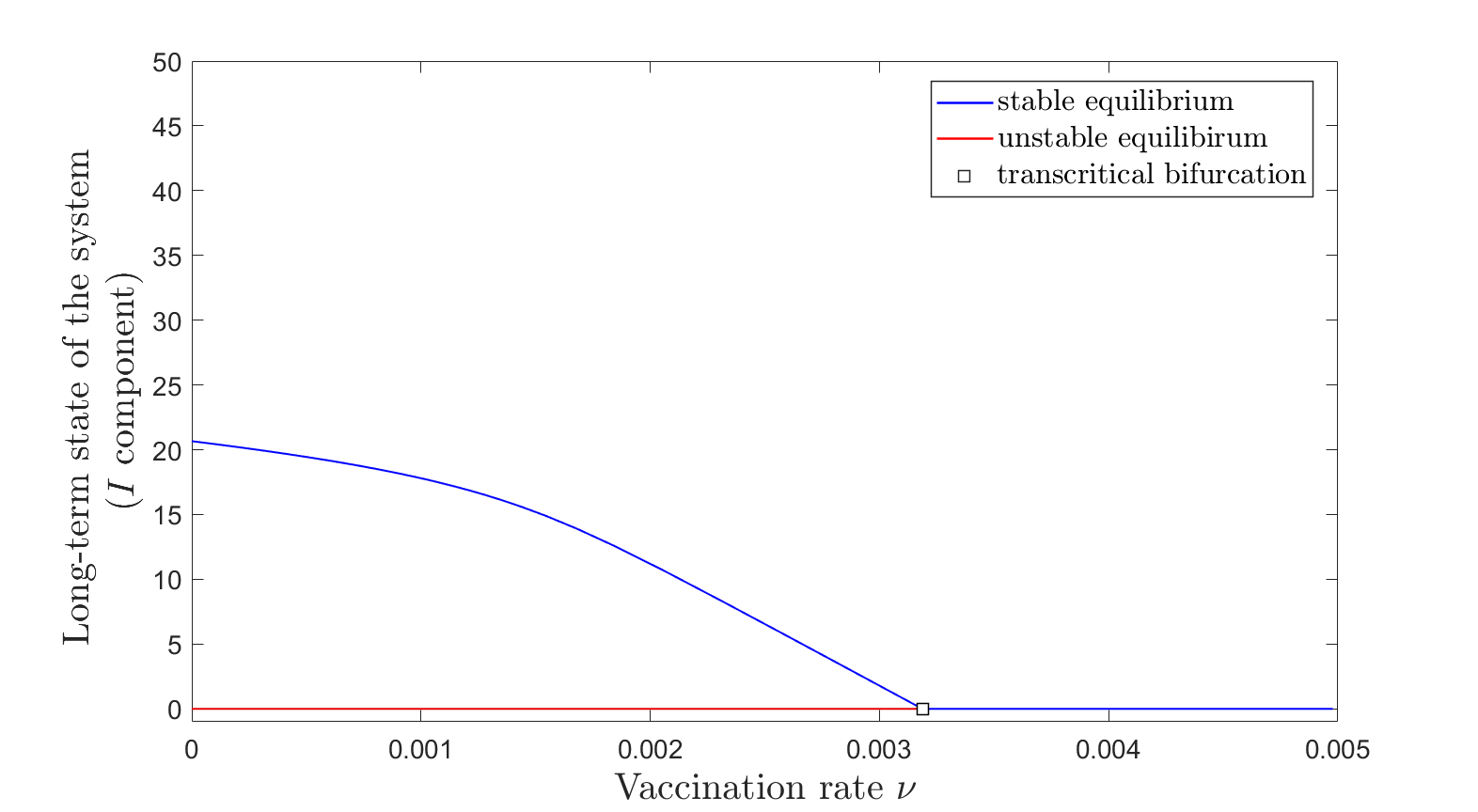}
\caption{\emph{{\bf Equilibria and bifurcations with respect to $\nu$, in the case of decreasing $\beta(I)$.} The $I^*$ component of the equilibirium branches is plotted in blue (if the equilibrium is stable) and in red (if the equilibrium is unstable). A transcritical bifurcation occurs at $\nu^*$ (marked on the graphs as a white square), where the stability of the equilibria ${\cal O}^*$ and ${\cal E}^*$ changes, as one real eigenvalue chages sign in each case. {\bf Left.} If the recovery period is $1/\gamma = 20$ days, $\nu^* \sim 0.004$. {\bf Right.} If the recovery period is $1/\gamma = 10$ days, $\nu^* \sim 0.0032$.}}
\label{decreasing_beta}
\end{center}
\end{figure}

The model predicts in this case that a vaccination rate as low as $\nu=0.004$ (equivalent roughly to five people out of 1000 being vaccinated every two days) is sufficient to insure clearance from an epidemic illness which takes $1/\gamma =10$ days for recovery. If the recovery time is $1/\gamma = 20$ days, the critical vaccination rate that guarantees clearance increases to $\nu^* = 0.0032$ (that is, about six people every two days). These are very low rates, even lower that in the similar case for scaled vaccination. This reinforces the idea that, in the scenario in which health protocols are scaled according to the Infection, vaccination protocols become another assisting tool, rather than the one critical mitigation strategy. That is because a important role in this mitigation is assumed by the health protocols. \\

\noindent \textbf{\emph{ Increasing vaccination when the infectiousness $\beta$ increases with case incidence.}} 

\noindent The sequence of transitions that occur in this case as the vaccination rate is decreased is similar to that observed for the shorter recovery rate, in the case of scaled vaccination: at first, the infection-free ${\cal O}^*$ is the only attractor, and guarantees epidemic clearance. As the vaccination crosses below the transcritical bifurcation, the system will converge to an endemic equilibrium with relatively low infection. After crossing a fold bifurcation, the system will also gain access to a stable cycle, for a wide interval of vaccination rates. Because ${\cal E}^*_1$ undergoes two supercritical Hopf bifurcations over this interval, losing and regaining stability, there will be ranges of this interval where the system has access both to the cycle ${\cal C}$ and the equilibrium ${\cal E}^*_1$ (depending on the state of the system), and ranges where the cycle is the only possible outcome. For longer recovery time, a second fold bifurcation eliminates the cycle ${\cal C}$, leaving a high endemic state as the only potential outcome for very low vaccination rates. Either way: large, persistent epidemic waves are a likely outcome, with the only alternative being an endemic steady state, with higher infection levels the lower the vaccination rate. However, these being said, note that these unwanted long-term prognoses occur for very small vaccination rates (bellow $\nu=0.004$ in both cases, equating to five people every two days).

\begin{figure}[h!]
\begin{center}
\includegraphics[width=0.48\textwidth]{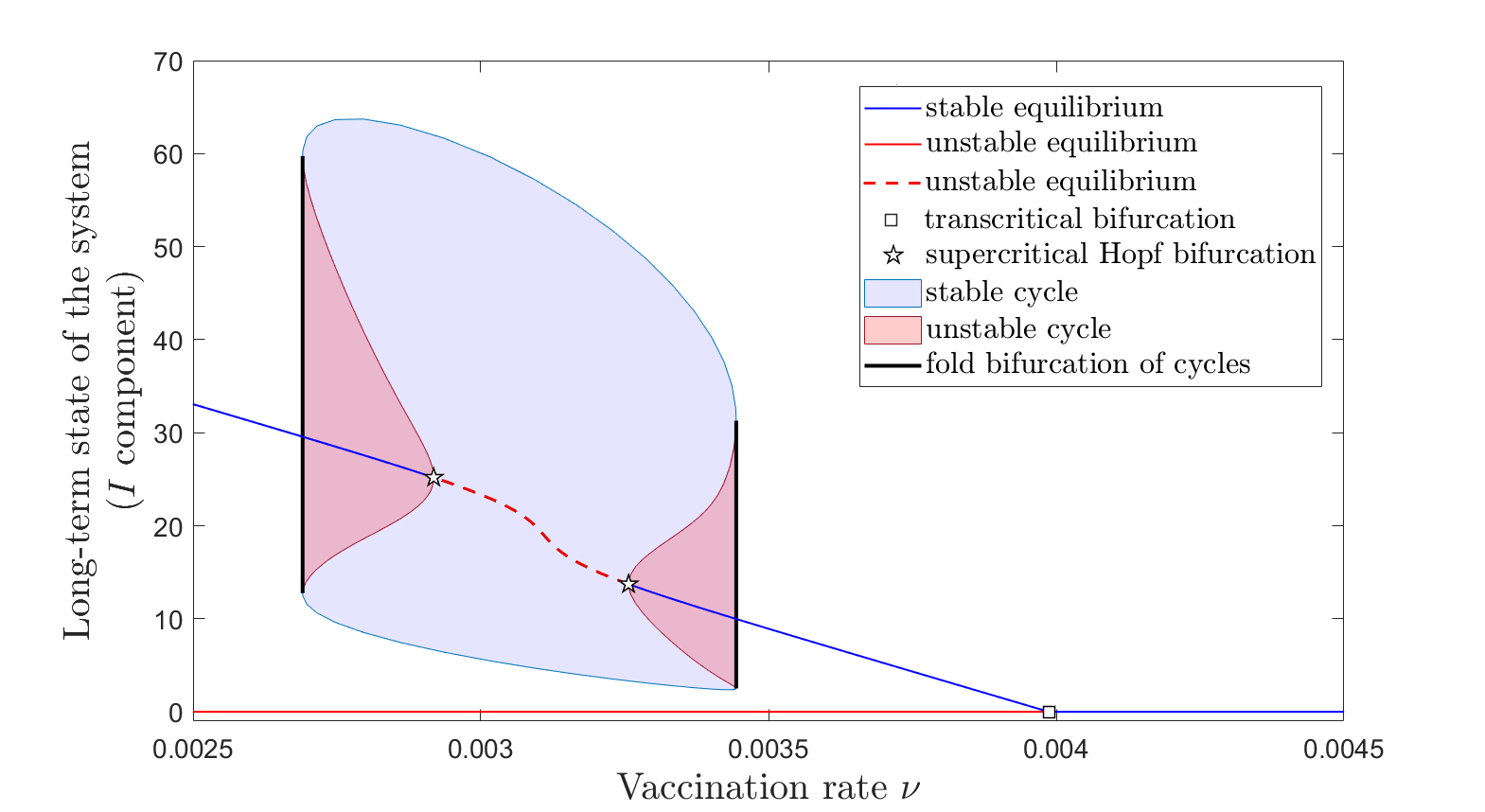}
\includegraphics[width=0.48\textwidth]{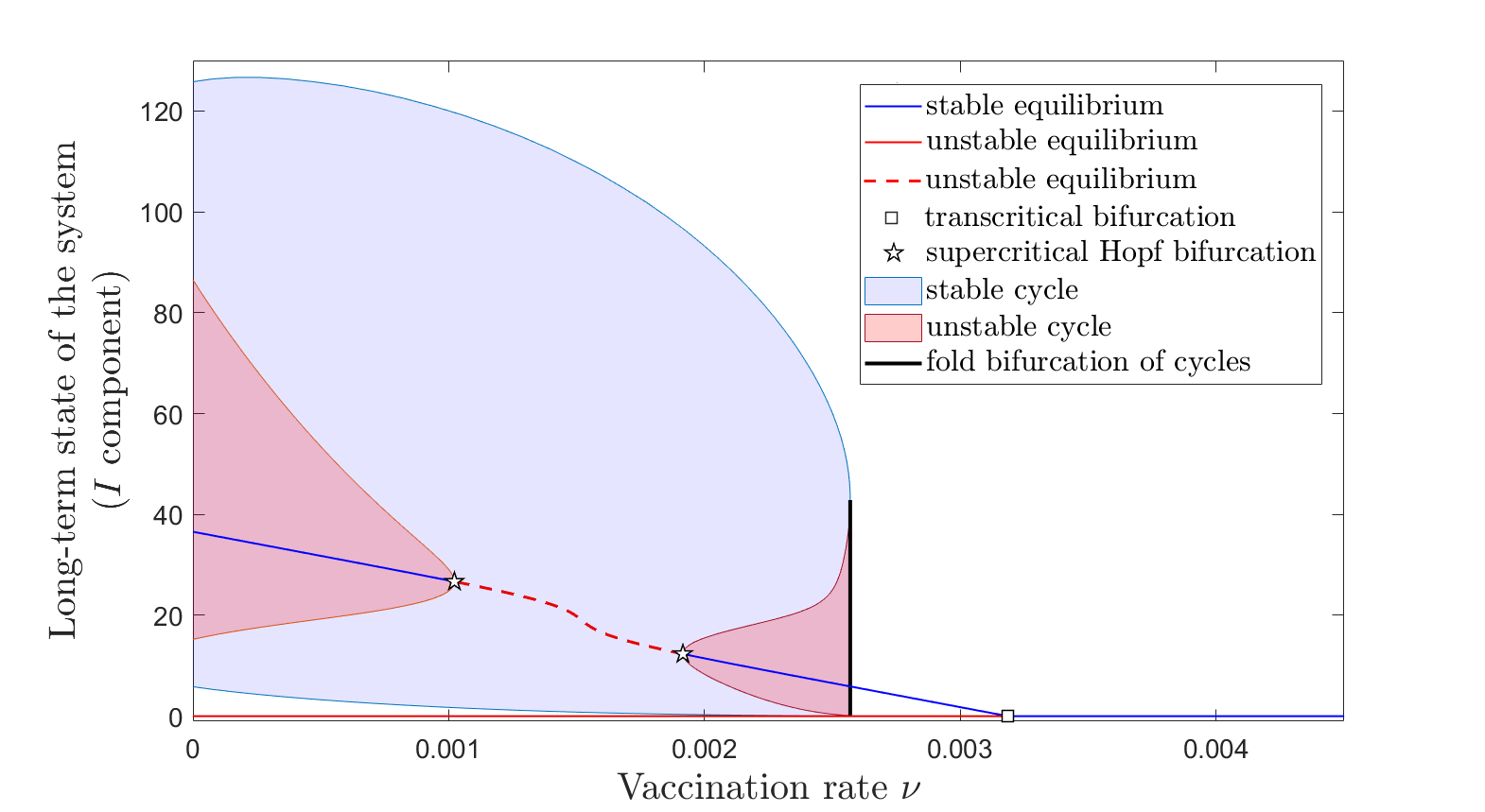}
\caption{\emph{{\bf Equilibria and bifurcations with respect to $\nu$, in the case of decreasing $\beta(I)$.} The $I$ component of the equilibirium branches is plotted in blue (if the equilibrium is stable) and in red (if the equilibrium is unstable). The formation and evolution of stable and unstable cycles are also shown, as their projection in the $I$ component. For a given $\nu$, the representation of the cycle is then a vertical segment, with its top and bottom (i.e., its min and max infection levels) evolving with $\nu$. Then the evolution of the cycle itself is shown as a shaded region between these bounding curves. Bifurcations are represented as follows: transcritical (white square); saddle node (white diamond); supercritical or subcritical Hopf (white star); fold, or saddle node of cycles (black bar). {\bf Left.} Bifurcation diagram for recovery period $1/\gamma = 20$ days. As vaccination is decreased from $\nu>0.004$ to $\nu=0$, the system transitions from convergence to an infection-free state, to two-state bistability, and eventually to a high-infection endemic state ($I^*\sim 80$ for $\nu=0$, not shown). {\bf Right.} Bifurcation diagram for recovery period $1/\gamma = 10$ days. As vaccination is decreased from $\nu>0.004$ to $\nu=0$, the system transitions, via a complex bifurcation sequence, from convergence to an infection-free state, to a low infection endemic state, then eventually to a bistability window in which an endemic equilibrium coexists with a stable cycle (epidemic waves).}}
\label{increasing_beta}
\end{center}
\end{figure}

First, we can notice again the crucial contribution of the recovery period to the overall dynamics and transitions. For example: in this scheme both lengths $1/\gamma = 10$ days and $1/\gamma =20$ days lead to endemic infection and the potential for epidemic waves for lack of compliance with health protocols and low vaccination rates. However, the endemic threshold $\beta^*$ occurs at a higher vaccination rate $\nu$ for the longer recovery period,  and the epidemic waves also appear at higher vaccination rates. At the vaccination rates where waves emerge in the 10-day recovery case, the oscillations have already ended in the 20-day recovery system, and the model predicts a high endemic state (with infection comparable with the peak of the oscillations in the 10-day case).

Second notice that, for this scheme as well, there is a qualitative difference between the effects of vaccination when the population is compliant with health protocols (decreasing $\beta(I)$) and when the population is acting against the common sense protocols (increasing $\beta(I)$). As before, the former leads to either complete viral clearance (when vaccination rates are sufficiently high), or to low endemic infection (20-50 cases) when vaccinations are very low. The latter leads to high endemic infection, or alternatively to wide annual epidemic waves, that can peak at 120 infections for very low vaccination rates. It clear that, in this scheme as well, adherence to health protocols delivers better mitigation results, even for very low vaccination rates. Next, we want to compare these results between the two vaccination schemes, and understand which one is more efficient in combination with adaptable population behavior, in what way to what extent.

\subsection{Comparison of behavioral effects between vaccination schemes}
\quad

\begin{figure}[h!]
\begin{center}
\includegraphics[width=0.48\textwidth]{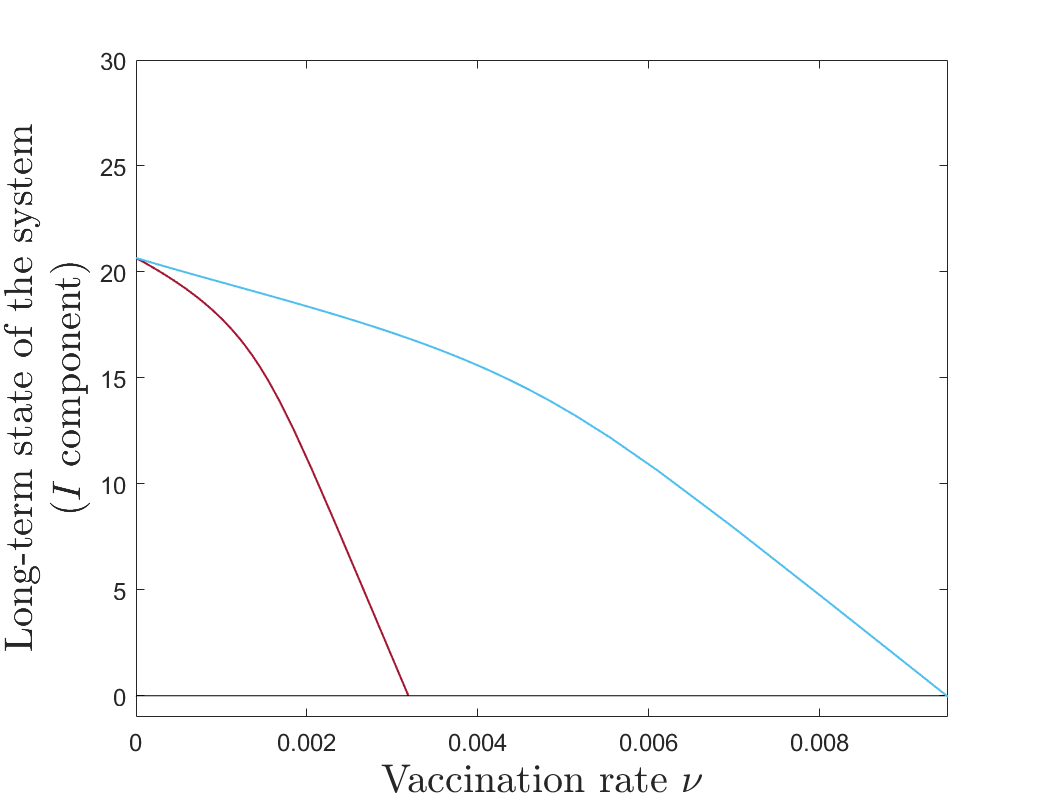}
\includegraphics[width=0.48\textwidth]{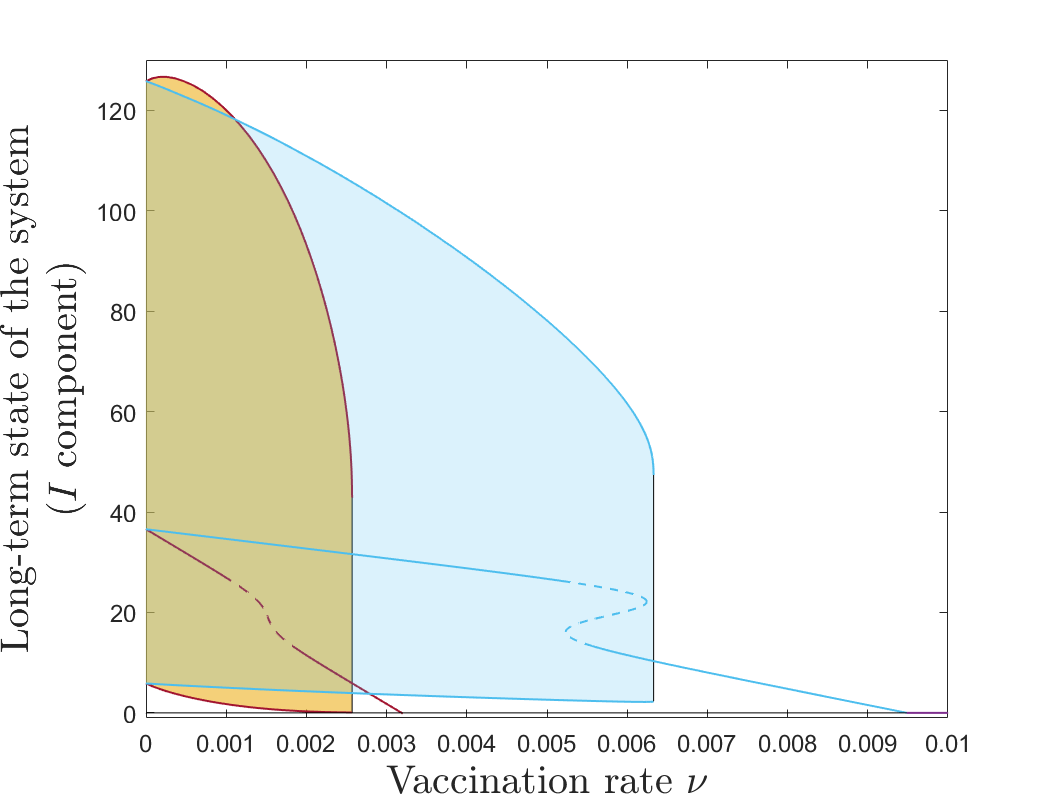}
\end{center}
\caption{\emph{{\bf Comparison between vaccination schemes.} {\bf Left.} The evolution of the endemic equilibrium is shown in terms of $\nu$, in brown (for constant vaccination) and in blue (for scaled vaccination). Fixed parameters: $\mu = 0.0008$, $\xi=0.004$ and $1/\gamma=10$. {\bf Right.} The evolution of the endemic equilibrium and the stable oscillations are shown in terms of $\nu$. For the constant vaccination scheme, the equilibrium and the top and bottom of the stable cycle are shown in brown; for the scaled vaccination scheme, they are shown in blue. Fixed parameters $\mu = 0.0008$, $\xi=0.004$ and $1/\gamma=20$. }}
\label{comparison_schemes}
\end{figure}

\vspace{2mm}
\noindent Based on the intuition from the previous section, comparing the two vaccination schemes for fixed infectiousness $\beta$, one could logically expect a similar effect here. And this intuition is generally correct: the constant vaccination scheme it more efficient in most ways than the scaled vaccination rate, in the sense that the former obtains more desirable results with the same rate $\nu$. The efficacy of each scheme can be quantified by the vaccination threshold $\nu^*$ where the endemic state takes over, or by the level of endemic infection for rates $\nu<\nu^*$. For both these quantifying measures, constant vaccination outperforms scaled vaccination. In all our simulations (for both typed of population behavior and for all sampled recovery rates), the endemic point occurs at a lower rate $\nu^*$, and the endemic infection $I_E^*$ is lower for the constant vaccination rate. This is briefly illustrated through two example in Figure~\ref{comparison_schemes}, for decreasing $\beta$ and $1\gamma=10$ days on the left, and for increasing $\beta$ and $1/\gamma=20$ on the right.

The constant rate scheme delivers a lower value for the transcritical bifurcation $\nu^*$, and lower endemic infection $I_E^*$ for all values of $\nu$ in the case of decreasing $\beta(I)$. Hence constant vaccination is decisively the better option when the population abides by health protocols.

These results remain true when $\beta(I)$ is increasing (i.e., the population behaves against health protocols). However, the behavior of the oscillatory regime does not fall under the same rule, as the model predicts that there are circumstances where the epidemic waves can in fact be more pronounced in the constant vaccination regime (see, for example, the low vaccination range in Figure~\ref{comparison_schemes}).

In conclusion, even though the comparison heavily favors the constant vaccination scheme, the fact that it does not outperform the scaled vaccination scheme across the board still leaves questions around which should be used, and under what circumstances. In fact, as mentioned before, it is likely that in reality these factors are a lot more fluid: the behavior of the population may switch in time between the two scenarios (working along versus against health protocols), and the vaccination scheme may itself change, for example due to availability, or compliance. Our analysis lays some basic results on which further work can develop, considering in more detail all these changing factors.

\section*{Acknowledgments}

The authors wish to thank: the Mathematical Association of America, National REU Program; the National Science Foundation, Award \#DMS-1950644; Dr. Nancy Campos and Jesslyn Burgos, SUNY New Paltz AC$^2$ Program, as well as the New Paltz Foundation.

\bibliographystyle{plain}
\bibliography{bibliography}

\end{document}